\documentstyle[12pt]{article}
0

\begin{document}

\begin{center}

~\vfill

{\Large \bf  The one-loop divergences of the linear gravity
with the torsion terms in tetrad approach}.

\vspace{1cm}

{\large M.~Yu.~Kalmykov}
\footnote { E-mail: $kalmykov@thsun1.jinr.dubna.su$ }

\vspace{2cm}

{\em Bogoliubov Laboratory of Theoretical Physics,
 Joint Institute for Nuclear  Research,
 $141~980$ Dubna $($Moscow Region$)$, Russian Federation}

\vspace{1cm}

{\large P.~I.~Pronin}\footnote{E-mail: $petr@theor.phys.msu.su$}

\vspace{1cm}
{\em Department of Theoretical Physics, Physics Faculty\\
Moscow State University, $117234$, Moscow, Russian  Federation}

\end{center}

\vfill

\begin{abstract}
In this paper we discuss the connection between the geometric and
tetrad approaches in the quantum affine-metric gravity. The
corresponding transition formulas are obtained at the one-loop level.
As an example, the one-loop counterterms are calculated in the tetrad
formalism in the theory with terms quadratic in the torsion field.
This model possesses the extra local symmetries connected with
transformation of the connection field. It is shown that the special
gauge can be chosen so that the corresponding additional ghosts do
not contribute to the one-loop divergent terms.
\end{abstract}

\vfill

PACS number(s) 0450+h, 0460-m

\pagebreak

\section{Introduction}

At the present moment there exists no perturbative renormalizable and
unitary quantum gravity. All suggested metric models of gravity are
nonrenormalizable \cite{R1}, \cite{R1a} or non-unitary \cite{R2}
ones.  The known models of the $N = 1$ supergravity are finite up to
two loops but may generate nonvanishing three-loop divergent
counterterms. Models with the extended $(e.g. N = 8)$ supersymmetry
or some other additional symmetry (e.g. the local conformal symmetry)
have better renormalization features, but there is no proof of their
complete finiteness by now.

The new type of quantum gravity is connected with a possibility for
the quantum (and possible classical) treatment of space-time to
involve more than the Riemannian space-time. The most interesting
non-Riemannian manifolds are the space-time with torsion \cite{FWH1}
and affine-metric space-time \cite{Hehl2}, \cite{Hehl1a}.  In these
geometries, there are geometric objects, additional to the metric
tensor, such as torsion and nonmetricity tensors defined as
independent variables. At the present moment there are a lot of
papers concerning the classical problems of the affine-metric
gravity, however, the renormalizability properties of the theory have
been studied insufficiently \cite{Neeman}. In particular, these
theories possess an additional symmetries connected with the local
transformation of the connection fields \cite{symmetries1}.
Additional global or local symmetries that are maintained at the
quantum level without generating anomalies may essentially improve
renormalization properties. The influence of these symmetries on the
renormalizable properties of the affine-metric quantum gravity is
discussed in papers \cite{MKL1}.  New hopes for a more perfect
quantum gravity arose in connection with string. The discussion of
the  bosonic string on the affine-metric manifold is given in paper
\cite{Vasiliy}.

The affine-metric manifold permits the geometric and tetrad description.
The geometric approach implies the description in terms of the metric
$g_{\mu \nu}$ and affine connection $\bar{\Gamma}^\sigma_{~\mu \nu}$.
The basic objects are expressed as:

\begin{itemize}
\item
curvature
$$
\bar{R}^\sigma_{~\lambda \mu \nu } (\bar{\Gamma}) =
\partial_\mu \bar{\Gamma}^\sigma_{~\lambda \nu}
 -
\partial_\nu \bar{\Gamma}^\sigma_{~\lambda \mu}
+ \bar{\Gamma}^\sigma_{~\alpha \mu} \bar{\Gamma}^\alpha_{~\lambda \nu}
- \bar{\Gamma}^\sigma_{~\alpha \nu} \bar{\Gamma}^\alpha_{~\lambda \mu}
$$
\item
torsion
$$
\bar{Q}^\sigma_{~\mu \nu }(\bar{\Gamma}) = \frac{1}{2}
\left( \bar \Gamma^\sigma_{~ \mu \nu} - \bar{\Gamma}^\sigma_{~\nu \mu }
\right)
$$
\item
nonmetricity
$$
\bar{W}_{\sigma \mu \nu}(g, \bar{\Gamma})
= \bar{\nabla}_\sigma g_{\mu \nu } = \partial_\sigma g_{\mu \nu }
- \bar{\Gamma}^\alpha_{~\mu \sigma} g_{\alpha \nu}
- \bar{\Gamma}^\alpha_{~\nu \sigma} g_{\alpha \mu}
$$
\end{itemize}

In the tetrad formalism, for describing the manifold we use the tetrad
$e^a_{~\mu}$  and the local Lorentz connection $\bar{\Omega}^a_{~b \mu}$.
Using the following relations \cite{Hehl1a}

\begin{eqnarray}
g_{\mu \nu} & = & e^a_{~\mu} e^b_{~\nu} \eta_{a b}
\label{connection1} \\
\bar{\nabla}_\sigma e^a_{~\mu} & = & \partial_\sigma e^a_{~\mu} +
\bar{\Omega}^a_{~b \sigma} e^b_{~\mu} - \bar{\Gamma}^\nu_{~\mu \sigma}
e^a_{~\nu} = 0
\label{connection2}
\end{eqnarray}
\noindent
where $\eta_{a b}$ is the Minkowskian metric,
we can obtain main geometric objects in the tetrad formalism

\begin{itemize}
\item
curvature
$$
\bar{R}^\sigma_{~\lambda \mu \nu } (\bar{\Gamma}) =
\bar{R}^a_{~b \mu \nu } (\bar{\Omega})  e_a^{~\sigma} e^b_{~\lambda} =
( \partial_\mu \bar{\Omega}^a_{~b \nu}
 -
\partial_\nu \bar{\Omega}^a_{~b \mu}
+ \bar{\Omega}^a_{~\alpha \mu} \bar{\Omega}^\alpha_{~b \nu}
- \bar{\Omega}^a_{~\alpha \nu} \bar{\Omega}^\alpha_{~b \mu} )
e_a^{~\sigma} e^b_{~\lambda}
$$
\item
torsion
$$
\bar{Q}^\sigma_{~\mu \nu }(\bar{\Gamma}) =
\bar{Q}^a_{~\mu \nu }(e,\bar{\Omega}) e_a^{~\sigma} =
-\frac{1}{2}
\left(
\partial_\mu e^a_{~\nu}  - \partial_\nu e^a_{~\mu}  +
\bar{\Omega}^a_{~b \mu} e^b_{~\nu}  -
\bar{\Omega}^a_{~b \nu} e^b_{~\mu} \right) e_a^{~\sigma}
$$
\item
nonmetricity
$$
\bar{W}_{\sigma \mu \nu}(g, \bar{\Gamma}) =
\bar{W}_{\sigma a b }(\bar{\Omega}) e^a_{\mu} e^b_{\nu}
= - \left( \bar{\Omega}_{ab \sigma} + \bar{\Omega}_{ba \sigma} \right)
e^a_{\mu} e^b_{\nu}
$$
\end{itemize}

The Lagrangian of a gauge theory is built out of terms quadratic in
the strength tensor of fields. The curvature is the field-strength
tensor of affine and local Lorentz connections. The field-strength
tensor for the metric and the tetrad is different. For the former it
is the nonmetricity tensor; for the later the torsion tensor. As
consequence, the Lagrangian of the affine-metric gravity will have the
different form in these two approaches.  Nevertheless, this
Lagrangian contains about two hundred of arbitrary parameters.  It is
very difficult to work with such a cumbersome expression.  Moreover,
some of these coefficients may be equal to zero. We don't know  what
coefficients are nonzero and only the construction of renormalizable
a theory of gravity may answer this question.  Hence, any Lagrangians
with dimension $ 2 $ and $ 4 $ terms constructed from the contraction
of the curvature, torsion and nonmetricity tensors may be considered
as a model of quantum affine-metric gravity.

The number and type of propagating fields depend on the choice of the
initial Lagrangian and description approach.  In the Lagrangian of
the affine-metric quantum gravity fields may exist which are
nonpropagating, nondynamical in one formalism (these fields are the
nonlinear second class constraints) and propagating, dynamical in
another.  Due to the unresolved problems of the loop calculations in
the theory with second class constraints it is desirable to use the
formalism in which all fields are propagating ones. Hence, one needs
to have the corresponding quantum transition expressions in order to
pass from one formalism to another.

In this paper we will obtain the transition expression
(\ref{quantum}) from one formalism to another at the quantum level in
the affine-metric gravity. As its application we calculate the
one-loop counterterms in the theory with terms quadratic in the
torsion field in the tetrad formalism.  This model possesses the
extra local symmetries connected with transformation of the
connection field.  We will research the influence of this symmetries
on the one-loop counterterms.

The following notation and conventions are accepted:

$$ c = \hbar = 1;~~~~~ \mu, \nu  = 0,1,2,3; ~~~~~a, b = 0,1,2,3;
~~~~~{\it k}^2 = 16 \pi G; ~~~~~\varepsilon  = \frac{4-d}{2}; $$

$$ e = det(e); ~~~~~(g) = det(g_{\mu \nu }), $$

Objects marked by bar are constructed by means of the affine connection
$\bar \Gamma^\sigma_{~\mu \nu}$. The others are the Riemannian objects.
Parentheses around index pairs denote symmetrization. The Riemannian
connection is
$\Gamma^\sigma_{~\mu \nu } = g^{\sigma \lambda} \frac{1}{2}
\left(- \partial_\lambda g_{\mu \nu} + \partial_\mu g_{\lambda \nu}
+ \partial_\nu g_{\lambda \mu} \right)$.
For further calculations one needs to define the following tensor object:
$ D^\sigma_{~\mu \nu }  = \bar{\Gamma}^\sigma_{~\mu \nu }
- \Gamma^\sigma_{~\mu \nu }$

\section{Connection between the geometric and tetrad approaches at the
one-loop level in the affine-metric quantum gravity}

To obtain the corresponding transition expressions at the one loop level
let us use the following method: introduce relations
(\ref{connection1}) and (\ref{connection2}) in the initial Lagrangian with
the corresponding Lagrange multipliers \cite{Hehl1a}:
\begin{equation}
S_{tot}  = S_{gr}(g,\bar{\Gamma},e,\bar{\Omega})  +
\frac{N^{(\mu \nu )}}{{\it k}^2}
\left( g_{\mu \nu } - e^a_{~\mu } e^b_{~\nu } \eta_{a b } \right)
+ \frac{M^{\sigma ~ \mu}_{~a}}{{\it k}^2} \bar{\nabla}_\sigma e^a_{~\mu }
\label{tot}
\end{equation}
where $S_{gr}(g,\bar{\Gamma},e,\bar{\Omega})$ may depend on all variables.
In this approach relations (\ref{connection1}) and (\ref{connection2})
are the second class constraints. The Lagrange multiplies are dynamical
variables. As a consequence, the Lagrange multiplies may be modified due to
loop corrections.

For obtaining the one loop transition expressions we will use the background
field method \cite{BDW}. In accordance with the background field method,
all dynamical variables are rewritten as a sum of classical and quantum parts:

\begin{eqnarray}
\underline{e}^a_{~\mu}  & = & e^a_{~\mu}  + {\it k} H^a_{~\mu },
~~~~~
\underline{\bar{\Omega}}^a_{~b \mu}   =  \bar{\Omega}^a_{~b \mu}
+ {\it k} \omega^a_{~b \mu },
\nonumber \\
\underline{\bar{\Gamma}}^\sigma_{~\mu \nu } & = &
\bar{\Gamma}^\sigma_{\mu \nu } + {\it k} \gamma^\sigma_{~\mu \nu },
~~~~~
\underline{g}_{\mu \nu }  =  g_{\mu \nu } + {\it
k}h_{\mu \nu }
\nonumber \\
\underline N^{\mu \nu}  & = & N^{\mu \nu }  + {\it k} n^{\mu \nu},
~~~~~
\underline M^{\mu ~\nu}_{~a}  =  M^{\mu ~\nu }_{~a}
+ {\it k} m^{\mu ~\nu}_{~a}.
\label{firstexpansion}
\end{eqnarray}

\noindent
where $g_{\mu \nu }, \bar{\Gamma}^\sigma_{~\mu \nu }, e^a_{~\lambda},
\bar{\Omega}^a_{~b \beta}, N^{\mu \nu }$ and $ M^{\sigma~\mu}_{~a}$
are classical parts satisfying the following equation of motion:

\begin{eqnarray}
\frac{\delta {\it S}_{tot}}{\delta \bar{\Omega}^a_{~b \sigma}} & = &
\frac{\delta {\it S}}{\delta \bar{\Omega}^a_{~b \sigma}} +
M^{\sigma ~ b}_{~a} = 0,
~~~~~~~~~~~~~~~~~~~~~~~~
\frac{\delta {\it S}_{tot}}{\delta \bar{\Gamma}^\nu_{~\mu \sigma}}  =
\frac{\delta {\it S}}{\delta \bar{\Gamma}^\nu_{~\mu \sigma}} -
M^{\sigma ~ \mu}_{~\nu} = 0
\nonumber \\
\frac{\delta {\it S}_{tot}}{\delta e^c_{~\sigma}} & = &
\frac{\delta {\it S}}{\delta e^c_{~\sigma}} - 2 N^{(\lambda \sigma)}
e_{c \lambda}
- \bar{\nabla}_\lambda M^{\lambda ~ \sigma}_{~c}
= 0,
~~~~~
\frac{\delta {\it S}_{tot}}{\delta g_{\mu \nu }}  =
\frac{\delta {\it S}}{\delta g_{\mu \nu }} + 2 N^{(\mu \nu)} = 0.
\label{motion}
\end{eqnarray}
where $N^{\mu \nu }$ and $M^{\sigma~ \lambda}_{~a}$ are the tensor densities.

Expanding the action ${\it S}_{tot}$ in powers of quantum fields up to the
terms quadratic in the quantum fields, we obtain the action for the
calculation of the one-loop counterterms

\begin{eqnarray}
{\it S}^{(2)}_{tot} & = &
n^{\mu \nu}(h_{\mu \nu } - H_{\mu \nu } - H_{\nu \mu })
- N^{\mu \nu } H^a_{~\mu } H_{a \nu}
+ M^{\sigma \mu}_{~a }(\omega^a_{~b \sigma} H^b_{~\mu}
-  \gamma^\nu_{~\mu \sigma } H^a_{~\nu})
\nonumber \\
& + & m^{\sigma ~ \mu}_{~a} (\bar{\nabla}_\sigma H^a_{~\mu}
+ \omega^a_{~\mu \sigma} - \gamma^a_{~\mu \sigma}) + {\it S}_{eff}
\nonumber
\end{eqnarray}

\noindent
where ${\it S}_{eff}$ is the action
${\it S}(g, \bar{\Gamma}, e, \bar{\Omega})$
quadratic in the quantum fields.

The action ${\it S}^{(2)}_{tot}$ is invariant under the general coordinate
transformations

\begin{eqnarray}
x^\mu & \rightarrow & 'x^\mu   =  x^\mu + {\it k} \xi^\mu(x)
\nonumber \\
H^a_{~\mu }(x) & \rightarrow & 'H^a_{~\mu }(x) =
- \partial_\mu \xi^\nu e^a_{~\nu }(x)
-  \xi^\nu \partial_\nu e^a_{~\mu }(x) + O({\it k})
\nonumber \\
w^a_{~b \mu }(x) & \rightarrow & 'w^a_{~b \mu }(x) =
-  \partial_\mu \xi^\nu \bar{\Omega}^a_{~b \nu }(x)
- \xi^\nu \partial_\nu \bar{\Omega}^a_{~b \mu }(x) + O({\it k})
\nonumber \\
h_{\mu \nu } & \rightarrow & 'h_{\mu \nu}  = h_{\mu \nu }
- \nabla_\mu \xi_\nu - \nabla_\nu \xi_\mu + O({\it k})
\nonumber \\
{\gamma}^\sigma_{~\mu \nu } & \rightarrow &
'{\gamma}^\sigma_{~\mu \nu } =
 \partial_\alpha \xi^\sigma \bar{\Gamma}^\alpha_{~\mu \nu }
- \partial_\mu \xi^\alpha \bar{\Gamma}^\sigma_{~\alpha \nu }
- \partial_\nu \xi^\alpha \bar{\Gamma}^\sigma_{~\mu \alpha }
- \partial_{\mu \nu } \xi^\sigma + O({\it k})
\nonumber
\end{eqnarray}

and under the local Lorentz rotations

\begin{eqnarray}
x^\mu & \rightarrow & 'x^\mu   =  x^\mu
+ {\it k} \Theta^\mu_{~\nu }(x) x^\nu
\nonumber \\
w^a_{~b \mu }(x) & \rightarrow & 'w^a_{~b \mu }(x) =
\Theta^a_{~c}  \bar{\Omega}^c_{~b \nu }(x)
-  \Theta^c_{~b}  \bar{\Omega}^a_{~c \mu }(x)
- \partial_\mu \Theta^a_{~b} + O({\it k})
\nonumber \\
H^a_{~\mu }(x) & \rightarrow & 'H^a_{~\mu }(x) =
\Theta^a_{~b} e^b_{~\mu }(x) + O({\it k})
\nonumber
\end{eqnarray}

The local Lorentz symmetry is fixed by means of the following
gauge \cite{R1a}:

\begin{equation}
H_{[\mu \nu ]}  = 0
\label{Landau}
\end{equation}

To violate the general coordinate transformation, we use the following gauge
condition:

\begin{eqnarray}
f^\mu & = & T^{\mu \sigma (\alpha \beta)} \nabla_\sigma h_{\alpha \beta}
+ (E^{\mu \rho \lambda~  \alpha \beta}_{~~~~\sigma}
\nabla_\rho \nabla_\lambda + G^{\mu ~ \alpha \beta}_{~ \sigma})
\gamma^\sigma_{~ \alpha \beta}
+ K^{\mu \sigma (\alpha \beta)} \nabla_\sigma H_{(\alpha \beta)}
\nonumber \\
{\it L}_{gf} & = & \frac{1}{2 \zeta} f^\mu \chi_{\mu \nu } f^\nu
\nonumber
\end{eqnarray}
where $\chi_{\mu \nu }$ is the differential operator;
$T^{\mu \sigma (\alpha \beta) }, K^{\mu \sigma (\alpha \beta)}
E^{\mu \rho \lambda~  \alpha \beta}_{~~~~\sigma}$
and $G^{\mu ~ \alpha \beta}_{~ \sigma}$ are the tensors depending on the
background fields, $\zeta$ is the gauge parameter. The ghost action is

$$
{\it L}_{gh}  =
- \left( \bar{c}^\mu ~\bar{C}^{\sigma \lambda} \right)
\left( \begin{array}{cc}
\chi_{\mu \pi} \biggl(
\left( T^{\pi \gamma (\omega \rho)} + K^{\pi \gamma (\omega \rho)} \right)
\nabla_\gamma \nabla_\rho + J^{\pi \omega} \biggr)
&  0 \\ \frac{1}{2} \left( \delta^\omega_\lambda \nabla_\sigma
- \delta^\omega_\sigma \nabla_\lambda \right)
+ \Omega_{\sigma \lambda}^{~~ \omega} & g_{\sigma \alpha} g_{\lambda \beta}
\end{array} \right)
\left( \begin{array}{c}
c_\omega \\
{C^{\alpha \beta}}
\end{array} \right) e
$$
where
$\{\bar{c}^\mu, c^\nu \}$ and $\{\bar{C}^{\sigma \lambda },C^{\mu \nu}\}$
are the ghost fields connected with the general coordinate and local Lorentz
transformations and

\begin{eqnarray}
J^\mu_{~ \omega} & = &
(E^{\mu \rho \lambda~  \alpha \beta}_{~~~~\sigma}
\nabla_\rho \nabla_\lambda + G^{\mu ~ \alpha \beta}_{~ \sigma})
\Bigr(\delta^\sigma_\omega D^\tau_{~\alpha \beta} \nabla_\tau
 - D^\sigma_{~\omega \beta} \nabla_\alpha
 - D^\sigma_{~\alpha \omega} \nabla_\beta
 - \nabla_\omega D^\sigma_{~\alpha \beta}
\nonumber \\
 & - & \frac{1}{2}(\nabla_\alpha \nabla_\beta + \nabla_\beta \nabla_\alpha)
 \delta^\sigma_\omega
 + \frac{1}{2}
(R^\sigma_{~\alpha \beta \omega} + R^\sigma_{~\beta \alpha \omega})
\Bigl)
\nonumber
\end{eqnarray}

\noindent
The gauge fixing term with higher derivatives may break the
unitary of the theory. To avoid this problem we consider the case
$E^{\mu \rho \lambda~  \alpha \beta}_{~~~~\sigma} = 0 $.
Introducing the notations
$s_{\mu \nu } = \frac{1}{2} \left( H_{\mu \nu} + H_{\nu \mu }
\right)$ and
$ t_{\mu \nu } = \frac{1}{2} \left( H_{\mu \nu} - H_{\nu
\mu } \right)$ we have

\begin{eqnarray}
e^{i W} & = & \int
exp \Bigl( S_{eff}(g, \bar{\Gamma}, e , \bar{\Omega}) + S_{gh} + S_{gf}
+ M^{\sigma ~ \mu}_{~a} (\omega^{a~\sigma}_{~b} H^b_{~\mu}
- \gamma^\nu_{~\mu \sigma} H^a_{~\nu})
\nonumber \\
&& - N^{\mu \nu } H^a_{~\mu } H_{a \nu}
\Bigr)
dh_{\alpha \beta}  ~d \gamma^\sigma_{~\mu \nu }
~d s_{\mu \nu }~dt_{\alpha \beta} ~d \omega^a_{~b \mu}
~d \bar{c}^\mu ~d c^\nu ~d \bar{C}^{\sigma \lambda } ~d C^{\mu \nu}
\nonumber \\
&& \delta ( h_{\mu \nu } - 2 s_{\mu \nu }) \delta (t_{\alpha \beta})
\delta (\bar{\nabla_\sigma} s^a_{~\mu} + \omega^a_{~\mu \sigma}
- \gamma^a_{~\mu \sigma}) ({\it det} \chi_{\mu \nu})^{ \frac{1}{2}}
\nonumber
\end{eqnarray}
Let us write the action  $S_{gr}(g, \bar{\Gamma}, e , \bar{\Omega}) $
of the affine-metric gravity either in the geometric formalism
$(g, \bar{\Gamma})$ or tetrad approach $(e, \bar{\Omega})$.
Then, from the equation of motion (\ref{motion}) we obtain

$$
M^{\sigma~\lambda}_{~a} = N^{\mu \nu } = 0
$$

The contribution of the Lorentz ghosts in the effective action is trivial in
the gauge (\ref{Landau}) in the dimensional regularization.
The one-loop generating functional is :

\begin{eqnarray}
e^{i W} & = & \int
exp \Bigl( S_{eff}(g, \bar{\Gamma}, e , \bar{\Omega})
+ \frac{1}{2 \zeta}  f^\mu \chi_{\mu \nu } f^\nu \Bigr)
dh_{\alpha \beta}  ~d \gamma^\sigma_{~\mu \nu }
~d s_{\mu \nu }~dt_{\alpha \beta} ~d \omega^a_{~b \mu}
~d \bar{c}^\mu ~d c^\nu
\nonumber \\
&& {\it det} \biggl( \chi_{\mu \pi} \left(
\left( T^{\pi \gamma (\omega \rho)} + K^{\pi \gamma (\omega \rho)} \right)
\nabla_\gamma \nabla_\omega + J^\pi_{~\rho} \right) \biggr)
\nonumber \\
&& \delta ( h_{\mu \nu } - 2 s_{\mu \nu }) \delta (t_{\alpha \beta})
\delta (\bar{\nabla_\sigma} s^a_{~\mu} + \omega^a_{~\mu \sigma}
- \gamma^a_{~\mu \sigma}) ({\it det} \chi_{\mu \nu})^{ \frac{1}{2}}
\nonumber
\end{eqnarray}

Hence, in the gauge (\ref{Landau}) the transition expressions from one
formalism to another are

\begin{eqnarray}
h_{\mu \nu }  & = & 2 H_{(\mu \nu) }
\nonumber \\
\gamma_{(\mu \nu) \sigma}  & = & \omega_{(\mu \nu) \sigma} +
\bar{\nabla}_\sigma H_{(\mu \nu) }
\nonumber \\
\gamma_{[\mu \nu] \sigma}  & = & \omega_{[\mu \nu] \sigma}
\nonumber \\
H_{[\mu \nu] } & = & 0
\label{quantum}
\end{eqnarray}

The contribution of the ghost fields to the one loop counterterms in
the gauge (\ref{Landau}) is independent of the choice of a formalism.

\section{The linear gravity with the torsion terms}

Consider a simple model with the terms quadratic in the
torsion fields. The Lagrangian of the model is the following:

\begin{equation}
{\it S}_{gr}  =  - \frac{1}{{\it k}^2} \int d^4 x \sqrt{-g}
\Bigl( \bar{R}(\bar{\Gamma} ) - 2 \Lambda
+ b_1 \bar{Q}_{\sigma \mu \nu} \bar{Q}^{\sigma \mu \nu}
+ b_2 \bar{Q}_{\sigma \mu \nu} \bar{Q}^{\nu \mu \sigma}
+ b_3 \bar{Q}_\sigma \bar{Q}^\sigma
\Bigr)
\label{action}
\end{equation}
where
$\Lambda $ is a cosmological constant, $\{b_i\}$ are
arbitrary constants and $\bar{Q}_\sigma = \bar{Q}^\lambda_{~\sigma \lambda}$

The torsion tensor has a different meaning within the geometric and
tetrad formalisms. In the tetrad approach the torsion tensor is the
strength tensor of tetrad fields, whereas in the geometric approach
the torsion tensor plays an auxiliary role. Let us calculate the
one-loop counterterms for the model (\ref{action}) within the tetrad
approach where $e^a_{~\mu}$ and $\bar{\Omega}^a_{~b \sigma}$ are
independent dynamical variables. For calculating the one-loop
counterterms we will use the background-field method \cite{BDW}
and expressions (\ref{quantum}).

In the case of special choice of the coefficients $\{b_j\}$ the
action (\ref{action}) is invariant under the extra local
transformations of the connection field \cite{symmetries1}.
This extra invariance is the sum of the projective invariance:

\begin{eqnarray}
\Gamma^\sigma _{~\mu \nu }(x) & \rightarrow  &
'\Gamma^\sigma _{~\mu \nu }(x)  = \Gamma^\sigma _{~\mu \nu }(x)
+ {\it k} \delta^\sigma_\mu C_\nu(x)
\label{projec}
\end{eqnarray}

\noindent
and antisymmetric one

\begin{eqnarray}
x^\mu & \rightarrow & 'x^\mu  = x^\mu \nonumber \\
g_{\mu  \nu }(x) & \rightarrow & 'g_{\mu \nu }(x)  = g_{\mu \nu }(x)
\nonumber \\
\Gamma^\sigma _{~\mu \nu }(x) & \rightarrow  &
'\Gamma^\sigma _{~\mu \nu }(x)  =
\Gamma^\sigma _{~\mu \nu }(x) + {\it k} g^{\sigma \lambda}
I_{[\lambda \mu \nu]}(x)
\label{additional}
\end{eqnarray}

\noindent
where $C_\nu (x)$ and $I_{\lambda \mu \nu }(x)$ are an arbitrary
vector and antisymmetric tensor respectively.
These invariances (\ref{projec}) and (\ref{additional}) arises from
the following choice of coefficients $\{b_j\}$:

$$
b_2 = b_1 - 1, ~~~~~
b_3 =  \frac{1}{3} - b_1
$$
The classical equations  of motion are:

\begin{eqnarray}
R_{\mu \nu} & = & \Lambda g_{\mu \nu}
\nonumber \\
D^\sigma_{~\mu \nu} & = & 0
\label{on-shell}
\end{eqnarray}

We will consider the theory with the additional invariances
(\ref{projec}) and (\ref{additional}).  For simplicity we will
calculate the one-loop counterterms on shell (\ref{on-shell}).

The one-loop Lagrangian on-shell quadratic in the quantum fields is

\begin{equation}
{\it L}_{eff} = - \frac{1}{2} \gamma^ \sigma_{~\mu \nu}
F_{~\sigma~~\lambda}^{~\mu \nu~\alpha \beta}
\gamma^\lambda_{~\alpha \beta}
  - \frac{1}{2} h^{\alpha \beta} ~X_{\alpha \beta \mu \nu} ~h^{\mu \nu}
-  \gamma^\lambda_{~\alpha \beta}  B_{\lambda ~~~\mu
\nu}^{~\alpha \beta \sigma} \nabla_\sigma h^{\mu \nu}
\label{expan}
\end{equation}
where

\begin{eqnarray}
F_{\alpha ~~\mu}^{~\beta \lambda ~ \nu \sigma} & = &
g^{\beta \lambda} \delta^\nu_\alpha \delta^\sigma_\mu
+ g^{\nu \sigma}\delta^\lambda_\alpha \delta^\beta_\mu
- b_1 g^{\beta \sigma} \delta^\nu_\alpha \delta^\lambda_\mu
- b_1 g^{\lambda \nu}\delta^\sigma_\alpha \delta^\beta_\mu
\nonumber \\
& + &  b_1~ g_{\alpha \mu} g^{\beta \nu} g^{\sigma \lambda}
- b_1~g_{\alpha \mu} g^{\lambda \nu} g^{\beta \sigma}
+ \frac{b_1-1}{2}
\left(
g^{\lambda \sigma} \delta^\beta_\mu \delta^\nu_\alpha
+ g^{\beta \nu}\delta^\lambda_\mu \delta^\sigma_\alpha
\right)
 \nonumber \\
& + &  \frac{1-3b_1}{6}
\left(
g^{\beta \nu}\delta^\lambda_\alpha \delta^\sigma_\mu
- g^{\beta \sigma }\delta^\nu_\mu\delta^\lambda_\alpha
- g^{\lambda \nu} \delta^\beta_\alpha\delta^\sigma_\mu
+ g^{\sigma \lambda} \delta^\beta_\alpha \delta^\nu_\mu
\right)
\nonumber
\end{eqnarray}

\begin{eqnarray}
P^{ \alpha \beta \mu \nu } & = & \frac{1}{4} \left(
g^{\alpha \mu } g^{\beta \nu } + g^{\alpha \nu }g^{\beta \mu } -
g^{\alpha \beta }g^{\mu \nu } \right)
\nonumber \\
B_{\lambda~~~\mu \nu}^{~\alpha \beta \sigma} & = &
2 \left(
\delta^\sigma_\lambda P^{\alpha \beta}_{~~ \mu \nu}
- \delta^\beta_\lambda P^{\alpha \sigma}_{~~\mu \nu} \right)
\nonumber \\
X_{\alpha \beta \mu \nu } & = & 2 \Lambda P_{\alpha \beta \mu \nu}
\nonumber
\end{eqnarray}

This expression was obtained in the geometric formalism.
We can rewrite the Lagrangian (\ref{expan})
in the tetrad formalism using expressions (\ref{quantum}).
For further calculations let us introduce the following notation:
$
\omega_{\sigma \mu \nu } = \omega_{[\sigma \mu]\nu}
+ \omega_{(\sigma \mu) \nu} \equiv
p_{[\sigma \mu]\nu} + l_{(\sigma \mu)\nu}.
$
Then,

\begin{eqnarray}
{\it L}_{eff} & = & \Biggl(
- \frac{1}{2} p^\sigma_{~\mu \nu}
F_{~~~\sigma~~\lambda}^{(3)~\mu \nu~\alpha \beta}
p^\lambda_{~\alpha \beta}
- 2 H^{\alpha \beta} H^{\mu \nu} X_{\alpha \beta \mu \nu}
 -  2 p^\lambda_{~\alpha \beta}
 B_{\lambda ~~~~\mu \nu}^{~\alpha \beta \sigma} \nabla_\sigma H^{\mu \nu}
\nonumber \\
&& - \frac{1}{2} \left( l^\sigma_{~\mu \nu} + \nabla_\nu
H^\sigma_{~\mu} \right) F_{~~~\sigma~~\lambda}^{(4)~\mu \nu~\alpha
\beta} \left( l^\lambda_{~\alpha \beta} + \nabla_\beta
H^\lambda_{~\alpha} \right)
\nonumber \\
&& - \left(
l^\lambda_{~\alpha \beta} + \nabla_\beta H^\lambda_{~\alpha} \right)
 2 B_{\lambda ~~~~\mu \nu}^{~\alpha \beta \sigma} \nabla_\sigma
 H^{\mu \nu} \Biggr) \sqrt{-g}
\label{eff1}
\end{eqnarray}

\noindent
where $\omega_{\sigma \mu \nu } $ and $ H_{\mu \nu }$  are defined in
(\ref{firstexpansion}) and

$$
F^{(3)\sigma \mu \nu \lambda \alpha \beta} =
F^{[\sigma \mu] \nu [\lambda \alpha] \beta},
~~~~~
F^{(4)\sigma \mu \nu \lambda \alpha \beta} =
F^{(\sigma \mu) \nu (\lambda \alpha) \beta}
$$

Expression (\ref{eff1}) is expansion of the initial Lagrangian
(\ref{action}) written in the tetrad formalism up to terms quadratic
in the quantum fields with the local Lorentz connection $\Omega^a_{~b
\sigma}$ and tetrad $e^a_{~\lambda}$ as independent dynamical
variables. In this expression the local Lorentz symmetry is broken by
the condition (\ref{Landau}). To define the propagator of the
quantum field $\omega^\sigma_{~\mu \nu} (\gamma^\sigma_{~\mu \nu})$
we must fix the additional symmetries (\ref{projec}) and
(\ref{additional}) at the quantum level.  The gauge conditions are

\begin{eqnarray}
f^{(1)}_\lambda & = &
A_1 \delta^\beta_\lambda \delta^\alpha_\sigma \gamma^\sigma_{~\alpha \beta}
= A_1 \delta^\beta_\lambda \delta^\alpha_\sigma
\left( \omega^\sigma_{~\alpha \beta} + \nabla_\beta H^\sigma_{~\alpha}
\right)
\label{pr_g}
\\
f^{(2)^\lambda} & = &
A_2 \varepsilon^{\lambda \sigma \alpha \beta} \gamma_{\sigma \alpha \beta}
= A_2 \varepsilon^{\lambda \sigma \alpha \beta} \omega_{\sigma \alpha \beta}
\label{ant_g}
\\
{\it L}^{add}_{gf} & = & \frac{1}{2} f^{(1) \lambda}f^{(1)}_\lambda
+ \frac{1}{2} f^{(2) \lambda}f^{(2)}_\lambda;
\nonumber
\end{eqnarray}

\noindent
where the constants $A_1$ and $A_2$ are nonzero.  We violate the
coordinate invariance of the action by means of the following gauge:

\begin{eqnarray}
F_\mu & = &
\nabla_\nu H^\nu_{~\mu} - \frac{1}{2} \nabla_\mu
H^\alpha_{~\alpha}
\nonumber
\\
{\it L}^{coor}_{gf} & = & 2 F_\mu F^\mu
\nonumber
\end{eqnarray}

The corresponding one-loop ghost on-shell Lagrangian is

\begin{equation}
{\it L}_{gh}  =
- \left( \bar{c}_\mu~\bar{C}_{\sigma \lambda}~\bar{\chi}_\mu ~\bar{\eta}_\mu
\right)
\left( \begin{array}{cccc}
\delta^\mu_\nu \left( \nabla^2 + \Lambda \right) & 0 & 0 & 0 \\
Z^{\sigma \lambda}_{~~~\nu}
& \delta^\sigma_\alpha \delta^\lambda_\beta & 0 & 0 \\
A_1 g^{\mu \omega} \nabla_\omega \nabla_\nu & 0 &
A_1 \delta^\mu_\nu & 0 \\
A_2 \triangle^\mu_{~\nu} & A_2 \epsilon^{\mu \omega \alpha \beta}
\nabla_\omega & 0 & \epsilon^{\mu \sigma \lambda \nu} \end{array}
\right)
\left( \begin{array}{c}
c^\nu \\
C^{\alpha \beta} \\
\chi^\nu \\
\eta_{\sigma \lambda \nu }
\end{array} \right)
\nonumber
\end{equation}

\noindent
where $\{\bar{c}_\nu, c^\mu,\} , \{ \bar{C}_{\alpha \beta}, C^{\mu \nu } \},
~\{ \bar{\chi}_\alpha, \chi^\beta \},
\{\bar{\eta}_\sigma, \eta^\lambda_{~\mu \nu } \}$
are anticommuting ghost fields connected with the general coordinate,
local Lorentz, projective and antisymmetric transformations, respectively.
$\triangle^\mu_{~\nu}$ and $Z^\mu_{~\nu}$ are

$$
Z_{\sigma \lambda }^{~~~\nu} =
\frac{1}{2} \left( \delta^\nu_\sigma \nabla_\lambda
- \delta^\nu_\lambda \nabla_\sigma  \right)
- \Omega_{\sigma \lambda}^{~~ \nu}
$$
$$
\triangle^\mu_{~\nu} = \epsilon^{\mu \lambda \alpha \beta}
\left( \Omega_{\alpha \beta \nu} \nabla_\lambda
+ \nabla_\nu \Omega_{\alpha \beta \lambda }
+ \Omega_{\alpha c \lambda} \Omega^c_{~\beta \nu }
- \Omega_{\alpha c \nu} \Omega^c_{~\beta \lambda}
\right)
$$

After inessential redefinition  of the ghost fields we obtain that
the one-loop contribution of the
$\{ \bar{C}_{\alpha \beta}, C^{\mu \nu } \},~
\{ \bar{\chi}_\alpha, \chi^\beta \},~
\{\bar{\eta}_\sigma, \eta^\lambda_{~\mu \nu } \}$
ghosts to the effective action is proportional to  $\delta^4(0)$. In
the dimensional regularization \cite{dim},
$[\delta^4(0)]_R = 0$ and the contribution of these ghosts to the
one-loop counterterms is equal to zero.

Replace the dynamical variables in the following way:

\begin{eqnarray}
\tilde p^\sigma_{~\mu \nu}  & = & p^\sigma_{~\mu \nu} +
 2 \overline{F}^{-1(3) \sigma~~ \lambda}_{~~~~~~~ \mu \nu~ \alpha
\beta} B^{~ \alpha \beta \tau}_{\lambda ~~~\rho \epsilon} \nabla _
 \tau  H^{\rho  \epsilon}
\nonumber \\
\tilde l^\sigma_{~\mu \nu}  & = & l^\sigma_{~\mu \nu}
+ \nabla_\nu H^\sigma_{~ \mu} + 2\overline{F}^{-1(4)
 \sigma~~ \lambda}_{~~~~~~~ \mu \nu~ \alpha \beta} B^{~ \alpha \beta
 \tau}_{\lambda ~~~\rho \epsilon} \nabla_\tau H^{\rho  \epsilon}
\label{replace}
\end{eqnarray}
where
$$
\overline{F}^{-1(3)\sigma \mu \nu \lambda \alpha \beta} =
\overline{F}^{-1 [\sigma \mu] \nu [\lambda \alpha] \beta},
~~~~~
\overline{F}^{-1(4)\sigma \mu \nu \lambda \alpha \beta} =
\overline{F}^{-1 (\sigma \mu) \nu (\lambda \alpha) \beta}
$$
\noindent
and

\begin{eqnarray}
\overline{F}^{-1 \alpha ~~\mu}_{~~~~\beta \sigma ~\nu\lambda}  =
 & - & \frac{1}{4}  g^{\alpha \mu} g_{\beta \sigma} g_{\nu \lambda}
 + \biggl( \frac{1}{4} + \frac{1}{9b_1} + \frac{1}{36 A_2} \biggr)
g^{\alpha \mu} g_{\beta \nu} g_{\sigma \lambda}
- \biggl( \frac{1}{12} + \frac{1}{18b_1} \biggr)
g_{\nu \beta} \delta^\mu_\lambda \delta^\alpha_\sigma
\nonumber \\
& + & \frac{1}{4}  \biggl( g_{\nu \lambda} \delta^\mu_\beta
\delta^\alpha_\sigma +
g_{\beta \sigma} \delta^\alpha_\nu \delta^\mu_\lambda \biggr)
- \biggl( \frac{1}{4} + \frac{1}{18b_1} - \frac{1}{36 A_2} \biggr)
\biggl( g_{\nu \sigma}
\delta^\alpha_\lambda \delta^\mu_\beta + g_{\beta \lambda}
\delta^\mu_\sigma \delta^\alpha_\nu \biggr)
\nonumber \\
& + &
\frac{1}{8}
\biggl( g_{\nu \lambda}
\delta^\mu_\sigma \delta^\alpha_\beta + g_{\beta \sigma}
\delta^\alpha_\lambda \delta^\mu_\nu \biggr)
+ \biggl( \frac{1}{4} - \frac{1}{9b_1} - \frac{1}{36 A_2} \biggr)
g^{\alpha \mu } g_{\sigma \nu } g_{\beta \lambda }
\nonumber \\
& - &
\biggl( \frac{1}{24} - \frac{1}{18b_1} \biggr)
\biggl( g_{\beta \lambda}
\delta^\mu_\nu \delta^\alpha_\sigma + g_{\sigma \nu} \delta^\alpha_\beta
\delta^\mu_\lambda \biggr)
 -
\biggl( \frac{1}{48} + \frac{1}{18b_1} - \frac{1}{16 A_1} \biggr)
g_{\sigma \lambda} \delta^\mu_\nu \delta^\alpha_\beta
\nonumber \\
& - &
\biggl( \frac{1}{4} - \frac{1}{18b_1} + \frac{1}{36 A_2} \biggr)
\biggl(
g_{\nu \beta } \delta^\mu_\sigma \delta^\alpha_\lambda  +
g_{\sigma \lambda } \delta^\mu_\beta \delta^\alpha_\nu  \biggr)
\label{prop}
\end{eqnarray}

\noindent
and  $b_1 \neq 0$.

The replacement (\ref{replace}) does not change the functional measure

$$ {\it det} \left|
\frac{\partial (H, ~\tilde{l}, ~\tilde{p}  )}{\partial (H,~l,~p)}
\right| = 1 $$

In the new variables the action (\ref{eff1}) is the following:

\begin{eqnarray}
{\it L}_{eff} & = &
\Biggl(- \frac{1}{2} \tilde p^ \sigma_{~\mu \nu}
\overline{F}_{~~~\sigma~~\lambda}^{(3)~\mu \nu~\alpha \beta}
\tilde p^\lambda_{~\alpha \beta}
- \frac{1}{2} \tilde l^ \sigma_{~\mu \nu}
\overline{F}_{~~~\sigma~~\lambda}^{(4)~\mu \nu~\alpha \beta}
\tilde l^\lambda_{~\alpha \beta}
- 2 H^{\mu \nu } H^{\sigma \lambda }
\left( R_{\mu \sigma \nu \lambda }  - R_{\mu \sigma } g_{\nu \lambda }
\right)
\nonumber \\
& + &  \nabla_\sigma H_{\alpha \beta}
\nabla_\lambda H^{\alpha \beta} g^{\sigma \lambda}
- 2 \nabla_\mu H^\mu_{~\nu} \nabla_\lambda H^{\lambda \nu}
+ 2  \nabla_\mu H^\sigma_{~\sigma} \nabla_\nu H^{\mu \nu}
-  \nabla_\sigma H^\mu_{~\mu } \nabla_\lambda H^\nu_{~\nu} g^{\sigma \lambda}
\nonumber \\
& - & 2 H^{\alpha \beta} H^{\mu \nu} X_{\alpha \beta \mu \nu}
\Biggr) e
\nonumber
\end{eqnarray}

\noindent
where
$
\overline F_{~~~\sigma ~~\lambda}^{~\alpha \beta ~\mu \nu}  =
F_{~~~\sigma ~~\lambda}^{~\alpha \beta ~\mu \nu} +
f^{(1) \lambda}f^{(1)}_\lambda + f^{(2) \lambda}f^{(2)}_\lambda
$.

The one-loop counterterms on the mass-shell including
the contributions of the quantum and ghost fields are

\begin{equation}
\triangle \Gamma^{(1)}_\infty = - \frac{1}{32 \pi^2 \varepsilon}
\int d^4x \sqrt{-g} \Biggl(
\frac{53}{45}
R_{\alpha \beta \mu \nu}R^{\alpha \beta \mu \nu}
 - \frac{58}{5} \Lambda^2 \Biggr)
\end{equation}

\section{Conclusions}

In the present paper we have obtained the transition expressions
(\ref{quantum}) which connect the geometric and tetrad
approaches at the one-loop level within the background formalism.
Using these expressions one-loop quantum corrections and
renormalization properties of the affine-metric quantum gravity can
be investigated in an arbitrary formalism. The essential step for
finding of these transition expressions is the gauge (\ref{Landau})
fixing the local Lorentz invariance.

Let us consider the model (\ref{action}) with extra local
symmetries (\ref{projec}) and (\ref{additional}).

\begin{enumerate}

\item
The projective (\ref{projec}) and antisymmetric (\ref{additional})
invariances do not influence on the renormalizability of the model
(\ref{action}).

\item
In the model (\ref{action}) the terms which are
quadratic over the torsion fields do not contribute to the one-loop
counterterms neither in the geometric nor in the tetrad approach.
So, we can assert, that in the theory with quadratic
in the torsion and nonmetricity Lagrangian these fields play an
auxiliary role at the quantum level, violating the
extra local symmetries of the affine-metric gravity.

\item
The special gauges ((\ref{pr_g}) and (\ref{ant_g})) are found so that
the corresponding action for ghosts related to the extra local
symmetries transformations has an algebraic form.  In any invariant
regularizaton with $[\delta(0)]_{ren} = 0$ its contribution
into one-loop counterterms is zero.

\item
The additional condition $(b_1 \neq 0)$ arising in the definition of
connection field propagator (\ref{prop}) corresponds to the new
symmetry of our theory.  We do not know how the connection field
transforms under this new symmetry. Only point we know exactly is
the metric and matter fields are invariant under these
transformations.

\item
The theory involved is renormalizable at the one-loop level on the
mass-shell.

\end{enumerate}


\begin{thebibliography}{200}

\bibitem{R1}
G. 't Hooft and M. Veltman,
Ann. Inst. H. Poincar\'{e} A {\bf 20} (1974) 69;

M. H. Goroff and A. Sagnotti, Nucl. Phys. B {\bf 266} (1986) 709.

\bibitem{R1a}
S. Deser and P. van Nieuwenhuizen, Phys. Rev. D {\bf 10} (1974) 411.

\bibitem{R2}
K. S. Stelle, Phys. Rev. D {\bf 16} (1977) 953.

\bibitem{FWH1}
F. W. Hehl,  P. van der Heyde,  G. D. Kerlick and I. M. Nester,
Rev. Mod. Phys. {\bf 48} (1976) 393;

K. Hayashi and T. Shirafyji,
Prog. Theor. Phys. {\bf 64} (1980) 866, 883, 1435, 2222;
ibid {\bf 65} (1981) 525;

F. W. Hehl, Found. Phys. {\bf 15} (1985) 451.

\bibitem{Hehl2}
L. Smolin, Nucl. Phys. B {\bf 247} (1984) 511;

J. Dell,  J. L. deLyra and L. Smolin,
Phys. Rev. D {\bf 34} (1986) 3012;

Y. Ne'eman and D.  \v{S}ija\v{c}ki,
Phys. Lett. B {\bf 200} (1988) 489;

F. W. Hehl,  J. D. McCrea,  E. W. Mielke and  Y. Ne'emann,
Found. Phys. {\bf 19} (1989) 1075.

\bibitem{Hehl1a}
F. W. Hehl,  J. D. McCrea,  E. W. Mielke and  Y. Ne'emann,
Phys. Rep. {\bf 258} (1995) 1 and References therein.

\bibitem{Neeman}
C. Y. Lee and Y. Ne'eman,
Phys. Lett. B {\bf 233} (1989) 286;
ibid, B {\bf 242} (1990) 59;

C. Y. Lee, Class. Quant. Grav. {\bf 9} (1992) 2001.

\bibitem{symmetries1}
V. D. Sandberg, Phys. Rev. D {\bf 12} (1975) 3013.

\bibitem{MKL1}
M. Yu. Kalmykov and  P. I. Pronin,
Nuovo Cimento B {\bf 106} (1991) 1401;
Gen. Rel. Grav. {\bf 27} (1995) 873;

M. Yu. Kalmykov,  P. I. Pronin and K. V. Stepanyantz,
Class. Quant. Grav. {\bf 11} (1994) 2645;

M. Yu. Kalmykov, Class. Quant. Grav. {\bf 14} (1997) 367.

\bibitem{Vasiliy}
V. E. Tarasov, Phys. Lett. B {\bf 323} (1994) 296.

\bibitem{BDW}
B. S. DeWitt,  Dynamical Theory Groups and Fields
(Gordon and Breach, New York, 1965).

\bibitem{dim}
G. 't Hooft and M. Veltman, Nucl. Phys. B {\bf 44} (1972) 189.

\end{thebibliography}
\end{document}